\documentclass[12pt,preprint]{aastex}
\usepackage{color}
\usepackage[normalem]{ulem}  % \sout{old text} for strikeout

\begin{document}

\title{A Comparative study of hyperon equations of state in supernova simulations}
\author{Prasanta Char\altaffilmark{1}}
\affil{Astroparticle Physics and Cosmology Division, Saha Institute of Nuclear 
Physics, 1/AF Bidhannagar, Kolkata-700064, India}
\altaffiltext{1}{Centre for Astroparticle Physics, Saha Institute of Nuclear
Physics, 1/AF Bidhannagar, Kolkata-700064, India}
\author{Sarmistha Banik}
\affil{BITS Pilani, Hyderabad Campus, Hyderabad-500078, India}
\author{Debades Bandyopadhyay\altaffilmark{1}} 
\affil{Astroparticle Physics and Cosmology Division, Saha Institute of Nuclear 
Physics, 1/AF Bidhannagar, Kolkata-700064, India}

\begin{abstract}
A comparative study of the $\Lambda$ hyperon equations of
state of Banik, Hempel and Banyopadhyay (BHB) \citep{bhb} and 
\citet{shen11} (denoted as HShen $\Lambda$) for 
core collapse supernova (CCSN) simulations is carried out in this work. 
The dynamical evolution of a protoneutron star (PNS) into a black hole
is investigated in core collapse supernova simulations in the general 
relativistic one dimensional code using the BHB$\Lambda \phi$ and 
HShen $\Lambda$ equation of state (EoS) tables and different progenitor models 
from Woosley and Heger \citep{Woos}. Radial profiles of the mass fractions of 
baryons, the density as well as the temperature in the PNS at different 
moments in
time, are compared for both EoS tables. The behaviour of the central density
of the PNS with 
time is demonstrated for those two $\Lambda$ hyperon EoS tables and compared
with their corresponding nuclear EoS tables. It is observed that the black hole 
formation time is higher in the BHB$\Lambda \phi$ case than in 
the HShen $\Lambda$ EoS for the entire set of progenitor models adopted here, 
because the repulsive $\Lambda$-$\Lambda$ interaction makes 
the BHB$\Lambda \phi$ EoS stiffer. Neutrino emission with the $\Lambda$ hyperon
EoS ceases earlier than that of its nuclear counterpart.    
The long duration evolution of the shock radius and gravitational mass of the
PNS after a successful supernova explosion 
with enhanced neutrino heating are studied with the BHB$\Lambda \phi$ EoS and 
$s$20WH07 progenitor model. The PNS is found to remain stable for 4 s and might
evolve into a cold neutron star.
\end{abstract}


\section{Introduction}
The recent discovery of a 2 M$_{\odot}$ neutron star puts stringent conditions 
on
the composition and equation of state (EoS) of dense matter in neutron star 
interiors \citep{anto}.
It has been observed that the presence of strangeness degrees of freedom such
as hyperons makes the EoS softer, which is incompatible with the massive neutron
star in most cases. This is known as the hyperon puzzle \citep{bub,lona}. 
Describing hyperon matter in neutron stars is a challenge in many-body 
theories. It has been argued that the hyperon-hyperon repulsive interaction due
to the exchange of strange vector meson makes the EoS stiffer and might 
overcome the puzzle.   

The $\beta$-equilibrated equations of state that include hyperons were 
constructed after the discovery of the massive neutron star by several groups. 
Those 
hyperon equations of state are found to result in 2 M$_{\odot}$ or heavier 
neutron stars \citep{weis1,weis2,last,colu,lopes,gus,vandalen14,char}. Besides
hyperons, the antikaon condensate was also included in some calculations, which
led to massive neutron stars \citep{char}. In all of these calculations, the 
repulsive hyperon-hyperon interaction that is mediated by $\phi$ mesons was 
considered.

Many EoS tables involving hyperons were developed for supernova simulations. 
The first hyperon EoS table was prepared by \citet{ishi}. In this case, the full
baryon octet was added to the Shen nucleon EoS table \citep{ishi,shen}. The Shen
nucleon EoS table was based on a relativistic mean field (RMF) model that had
the Thomas-Fermi approximation for the description of inhomogeneous matter below
the saturation density \citep{shen}. Later, the Shen nucleon EoS was extended to
include only $\Lambda$ hyperons in the HShen $\Lambda$ EoS table 
\citep{shen11}.
Another extensively used supernova EoS is the Lattimer and Swesty (LS) nucleon
EoS table, which based on the non-relativistic Skyrme interaction \citep{ls}. 
Recently $\Lambda$ hyperons were included in the LS nucleon EoS 
\citep{oertel12}. All these hyperon EoS tables were used in core collapse 
supernova (CCSN) simulations by several authors 
\citep{ishi,naka08,naka,sumi,sb1,sb2}. However, none of these hyperon EoS 
tables were consistent with the 2 M$_{\odot}$ neutron star constraint.

Recently, we computed EoS tables that included $\Lambda$ hyperons within the 
framework of the density dependent relativistic hadron (DDRH) field theory 
\citep{bhb}. In those EoS tables,
light and heavy nuclei, as well as interacting nucleons are described in 
the nuclear statistical equilibrium model which takes into account the excluded
volume effects \citep{hs1,bhb}. Two variants of the hyperon EoS tables were 
generated - 
in one case (BHB$\Lambda\phi$), the repulsive $\Lambda$ hyperon - $\Lambda$ 
hyperon interaction mediated by $\phi$ mesons was considered, and in the other 
case (BHB$\Lambda$), this interaction was neglected. It should
be noted that the DDRH model with the DD2 parameter set for nucleons is 
in very good agreement with the symmetry energy properties at the saturation
density \citep{typ10,composemanual,jim}. We imposed the charge 
neutrality and $\beta$-equilibrium conditions on the BHB hyperon EoS tables and
calculated the mass-radius relationship of the neutron star sequence. It was
observed that the maximum mass corresponding to the BHB$\Lambda\phi$ EoS was
2.1 M$_{\odot}$ which is well above the recently observed massive neutron star
\citep{bhb}. Other hyperon EoSs for $\beta$-equilibrated 
neutron star matter gave rise to the maximum mass neutron stars of 1.75 
M$_{\odot}$ for the HShen $\Lambda$ EoS \citep{shen11}, 1.6 M$_{\odot}$ 
for Ishizuka EoS \citep{ishi}, 1.91 M$_{\odot}$ for the LS+$\Lambda$ EoS 
\citep{oertel12}.  
    
In this paper, for the first time, we perform supernova simulations with 
the BHB$\Lambda\phi$ EoS 
table, which is compatible with a 2 M$_{\odot}$ neutron star, in the 
general relativistic one dimensional (GR1D) hydrodynamics code \citep{ott10}.
Our main goal is to investigate the appearance of $\Lambda$ hyperons in the 
postbounce phase and the role of $\Lambda$ hyperons in CCSNs. Furthermore, 
we compare our simulation results with those of previous calculations with other
hyperon EoS tables, particularly the HShen $\Lambda$ EoS table 
\citep{sb1}. We are looking for important effects of hyperons in CCSN with the 
BHB$\Lambda\phi$ EoS compared with those of other hyperon EoS. 

The paper is organised as follows. In Section 2, the DDRH model for
BHB$\Lambda\phi$ EoS table and the RMF model for the HShen $\Lambda$ EoS 
table are described. We also briefly discuss the GR1D model for CCSN 
simulations. The results of our calculation are discussed and compared with 
those 
of the HShen $\Lambda$ EoS in Section 3. Section 4 contains the summary and 
conclusions. 

\section{Methodology}
First 
First we shall discuss the salient feature of the BHB$\Lambda \phi$ and 
HShen $\Lambda$ EoS tables for CCSN simulations \citep{shen11,bhb}. The EoS 
tables are functions of three parameters
i.e. baryon number density, temperature, and proton fraction. In both cases, 
the compositions of matter that vary from one region to the other depending on
those parameters are nuclei, (anti)neutrons, (anti)protons, 
(anti)$\Lambda$ hyperons, photons plus electrons and positrons that form a 
uniform background. The contribution of (anti)neutrinos is not added to the EoS
tables and is dealt with separately. We describe the baryonic contribution 
below.

\subsection{BHB$\Lambda \phi$ and HShen $\Lambda$ EoS tables}

In the BHB$\Lambda \phi$ EoS table, the 
nuclear statistical equilibrium (NSE) model of \citet{hs1} is applied for the 
description of the matter made of light and heavy nuclei, and unbound nucleons 
at low temperatures and below the saturation density, whereas the high density 
matter is described within the framework of the DDRH model adopting the RMF 
approximation \citep{bani02,typ10,bhb}. The repulsive interaction between 
$\Lambda$ hyperons mediated by $\phi$ mesons is included in the RMF model.
Nucleon-meson couplings in the DDRH model are density dependent. The DD2 
parameter set of nucleon-meson couplings is used to describe the nuclear matter
properties \citep{wol,typ10,composemanual,fis2}. It should be noted that the 
nuclear EoS in the DDRH model using DD2 parameter set is known as HS(DD2) 
\citep{fis2}. 

On the other hand, the uniform matter at high density and temperature in 
the HShen $\Lambda$ EoS table was described within the framework of the RMF 
model including nonlinear terms in $\sigma$ and $\omega$ mesons 
\citep{shen11}; non-uniform matter 
at low temperatures and below the saturation density was considered as a 
mixture 
of alpha particles, heavy nuclei, and unbound nucleons. Heavy nuclei were 
calculated using the Thomas-Fermi approach. The Shen EoS exploited the single 
nucleus approximation for heavy nuclei \citep{shen,shen11}.
The interaction among $\Lambda$ hyperons due to $\phi$ mesons was neglected in 
this case. Furthermore, in this case baryon-meson couplings of the RMF model 
are density-independent. We 
denote the EoSs with and without $\Lambda$ hyperons as HShen $\Lambda$ and 
HShen, respectively. The parameter set from \citet{suga} that is known as the 
TM1 set was adopted for the nucleon-meson coupling constants of the RMF model. 

The nuclear matter saturation properties of two RMF models discussed above are
recorded in Table {\ref{table}}. It should be noted that though the 
incompressibility 
of nuclear matter, symmetry energy, and its slope coefficient of the DD2 set at 
the saturation density are in very good agreement with experimental values 
\citep{jim,fis2}, the corresponding quantities of the TM1 set are not. This 
would 
have serious bearing on the description of high density matter in the RMF model
of HShen \citep{shen11}. For both EoS tables, $\Lambda$ hyperon-vector meson 
couplings are estimated from the SU(6) symmetry relations \citep{dov,sch} and
$\Lambda$ hyperon - scalar meson coupling is obtained from the hypernuclei 
data. The $\Lambda$ hyperon potential depth is -30 MeV in normal nuclear 
matter \citep{mil,mar,sch92}.    

The EoSs of $\beta$-equilibrated and charge neutral cold neutron star matter 
with and without $\Lambda$ hyperons are calculated from the supernova 
EoS Tables. The maximum masses of cold neutron stars without $\Lambda$ hyperons
for HS(DD2) and HShen EoS are given by Table {\ref{table}}. 
Furthermore, the 
maximum masses of cold neutron stars corresponding to the BHB$\Lambda\phi$ and 
HShen $\Lambda$ are  2.1 M$_{\odot}$ and 1.75 M$_{\odot}$ \citep{bhb,shen11},
respectively.

For CCSN simulations, we make use of the 
HS(DD2), BHB$\Lambda \phi$, HShen and HShen $\Lambda$ EoS tables which are 
available from the stellarcollapse.org website
\footnote{See \texttt{http://stellarcollapse.org/equationofstate}}.

\subsection{General relativistic model for supernova simulations}
We perform the CCSN simulations using the spherically symmetric general
relativistic hydrodynamics code GR1D which was developed by \citet{ott10}. 
Microphysical EoSs for supernova matter and an approximate treatment of 
neutrinos
in the pre- and postbounce phases are implemented in the GR1D code. We use the
BHB$\Lambda \phi$ and HShen $\Lambda$ EoS tables in CCSN simulations with the 
GR1D code. Three neutrino species denoted by $\nu_e$, $\bar{\nu}_e$, 
$\nu_x (= \nu_{\mu}, {\bar{\nu}_{\mu}}, \nu_{\tau}, {\bar{\nu}_{\tau}})$ 
are considered in this model \citep{ott11}. Key aspects of neutrino heating and
cooling are incorporated into the model. The leakage scheme \citep{ruf,ros} 
exploited in the GR1D code gives approximate number and energy emission rates. 
The neutrino heating rate considered here involves the scale factor $f_{heat}$ 
which could be enhanced beyond the normal value of 1 to achieve additional 
neutrino heating for "successful" CCSN explosions \citep{jank,ott11}. We take 
$f_{heat} = 1$ in CCSN simulations, if not stated otherwise.

In principle, an accurate and expensive neutrino treatment should be based 
on 
the Boltzmann neutrino transport. However, computationally efficient schemes for
neutrinos are employed in the GR1D code for faster CCSN simulations. Moreover,
it has been noted that the results obtained in CCSN simulations using the 
simplified treatment of neutrino leakage and heating in the GR1D were  
quantitatively similar to the results obtained from one dimensional (1D) 
simulations with the 
Boltzmann neutrino transport by other groups \citep{fis09,sumi07}. It was 
argued that progenitor structures played more important roles in the collapse 
of a protoneutron star (PNS) to a black hole than the details of neutrino 
treatment \citep{ott11}. 

\section{Results and Discussion}       
Now we report our investigations on CCSNs within the GR1D code using the HShen 
$\Lambda$ hyperon and BHB$\Lambda \phi$ EoS tables. In these studies, 
nonrotating progenitors of
\citet{Woos} (WH07) are used. In their stellar evolution studies \citet{Woos} 
evolved zero age main-sequence (ZAMS) stars
with solar metallicity denoted by the prefix $s$ before presupernova models, 
followed by ZAMS mass. Significant mass loss was reported in $s$WH07 
presupernova models \citep{ott11}.

We perform the CCSN simulations with presupernova models as recorded in 
Table \ref{table1}. In all numerical calculations, we fix the neutrino 
heating
factor $f_{heat} = 1$. In the next paragraphs, we discuss the results of 
simulations starting from the gravitational collapse of the iron core followed 
by the core bounce to the postbounce evolution of the PNS 
for $s$40WH07 and $s$23WH07 models with the HShen $\Lambda$ and 
BHB$\Lambda \phi$ EoS tables in details. In all of these simulations, 
a shock wave is launched at the core bounce, it stalls after traversing
a few 100 km, then recedes and becomes an accretion shock. Because neutrinos in
the 1D CCSN model could not revive the shock, the PNS shrinks due
to mass accretion and its density and temperature increase during the 
postbounce evolution. This leads to the appearance of $\Lambda$ hyperons in the
PNS.        

For $s$40WH07, the core bounce occurs at 0.273 and 0.321 s, corresponding to
the HShen $\Lambda$ hyperon and BHB$\Lambda \phi$ EoS, respectively. 
Similarly, in the $s$23WH07 model the core bounce times for the HShen $\Lambda$ 
and the BHB$\Lambda \phi$ EoS are 0.266 and 0.315 s, respectively. 
The appearance of 
strangeness or $\Lambda$ hyperons in the postbounce phase and its role in the 
evolution of the PNS are the main focuses of this investigation. For $s$40WH07 
and
$s$23WH07 models and both hyperon EoS tables, $\Lambda$ hyperons do not 
populate the PNS at the core bounce. In simulations with both presupernova 
models, strangeness in the form of $\Lambda$ hyperons sets in a few hundred 
milliseconds (ms) after the core bounce and increases with time thereafter.

Figure 1 depicts the PNS compositions as a function of radius at two different 
postbounce times for $s$40WH07 with the HShen $\Lambda$ (left panel) and 
BHB$\Lambda \phi$ (right panel) EoS tables. For postbounce time ($t_{pb}$) 
0.31 s, the central value of $\Lambda$ fraction is higher for the 
BHB$\Lambda \phi$
EoS than that of the HShen $\Lambda$ hyperon EoS. The profile of $\Lambda$ 
hyperons is
wider in the latter case. We find similar trends for $\Lambda$ hyperons at a 
later time $t_{pb}=$ 0.51 s. For both EoS tables, the population of $\Lambda$s
increases with time. It is to be noted that the central value of the $\Lambda$ 
fraction is a high density effect, whereas the off-centre $\Lambda$s are 
populated thermally. We study the density and temperature profiles to 
understand this behaviour. 

The density profiles as a function of radius are plotted for $s$40HW07 at the 
bounce as well as for $t_{pb} = $ 0.31 and 0.51 s in Figure 2. The left panel 
of the figure corresponds to the HShen $\Lambda$ EoS and the right panel 
implies the results of the BHB$\Lambda \phi$ EoS. At the bounce, the central 
density ($\rho_c$) of the
PNS in both cases is just above the normal nuclear matter density, as evident 
by the figure. Though the density profiles for both EoS tables are
quantitatively the same at $t_{pb} =$ 0, they differ at later times. The 
central density at $t_{pb} =$ 0.51 in the right panel is higher than that of 
the left panel. In both cases, the central density exceeds two times the normal 
nuclear matter density. This high central density facilitates a significant 
population of $\Lambda$s in the core of the PNS, as seen in Fig. 1. However, the
density falls well below normal nuclear matter density at the tail of the 
profile. The off-centre $\Lambda$s in Fig. 1 could not be attributed to the 
density effect.

The temperature profiles as a function of radius are shown for $s$40WH07, with
the HShen $\Lambda$ hyperon (left panel) and the BHB$\Lambda \phi$ 
(right panel) 
EoS tables in Figure 3. Just as in Figure 2, the temperatures profiles are 
plotted at
the core bounce and $t_{pb} =$ 0.31 and 0.51 s in both panels of Figure 3. 
The peaks of temperature profiles located away from the centre of the PNS for 
both EoSs after the core bounce later shift toward the centre with time in both
panels. It is to be noted that the central temperature at the bounce is higher 
for the BHB$\Lambda \phi$ EoS compared with the corresponding temperature for
the HShen $\Lambda$ EoS. Furthermore, the peak temperature around 8 km at 
0.51 s after core bounce in the case of the BHB$\Lambda \phi$ EOS is much 
higher 
than the corresponding scenario for the HShen $\Lambda$ EoS. This high 
temperature results in thermally produced $\Lambda$ hyperons away from the 
centre of the PNS as shown in Figure 1. We find from Figure 1 that thermal 
$\Lambda$s are more abundant around 8 km at later times for the 
BHB$\Lambda \phi$ EoS due to a higher peak temperature.       

We also study profiles of particle fraction, density, and temperature for
$s$23WH07 using both hyperon EoS tables as shown in Figures 4-6. We obtain 
qualitatively similar results for $s$23WH07 as we have already discussed for 
$s$40WH07.  

Now we focus on the postbounce evolution of the PNS for different presupernova
models with nuclear and $\Lambda$ hyperon EoS tables corresponding to the HShen
and BHB models. Figure 7 exhibits 
the evolution of the central density of the PNS with the postbounce times for
$s$40WH07 (left panel) and $s$23WH07 (right panel). Results are shown in both
panels for the HShen nuclear EoS, the HShen $\Lambda$ EoS, 
the HS(DD2) nuclear EoS and the BHB$\Lambda \phi$ EoS. It should be noted that 
the
core bounce time for the hyperon EoS is the same as that of the corresponding 
nuclear
EoS. In all cases in both panels of Figure 7, we find that the central density 
increases gradually to several times the normal nuclear matter density.
Finally, there is a steep rise in the central density when the PNS dynamically
collapses into a black hole in milliseconds. It should be noted that the black 
hole 
formation time is different for different EoS models. It is evident from 
the CCSN simulation of $s$23WH07 that the black holes are formed at 1.511 and
1.623 s after the core bounce for the HS(DD2) and the HShen EoS, respectively.
For $s$40WH07, the black hole formation time is 0.942 s in the case of the 
HS(DD2) 
EoS, whereas it is 1.084 s for the HShen EoS. For both supernova models and 
nuclear EoS tables, the black hole is formed earlier in case of the HS(DD2) 
than the situation with the HShen EoS. The maximum gravitational (baryonic) PNS
masses are 2.464 (2.616) M$_{\odot}$ and 2.459 (2.587) M$_{\odot}$ for 
$s$40SW07 with 
the HS(DD2) and the HShen EoS, respectively. Similarly, for $s$23WH07, those 
are 
2.428 (2.605) M$_{\odot}$ and 2.383 (2.512) M$_{\odot}$ corresponding to the
HS(DD2) and the HShen EoS. On the other hand, the dynamical collapse to
a black hole is accelerated for the HShen $\Lambda$ and BHB$\Lambda \phi$
EoS tables because hyperons make the EoS softer. It is evident from Figure 7 
that
the black hole formation time is shorter for hyperon EoS than that for the 
corresponding nuclear EoS. However, there is little difference between the
black hole formation times corresponding to the HShen $\Lambda$ and 
BHB$\Lambda \phi$ EoSs.  

The results of CCSN simulations with other presupernova models are recorded in
Table \ref{table1}. The first column of the table lists the presupernova models
of
\citet{Woos} starting from $s$20WH07 to $s$40WH07. Two EoS tables, such as 
the HShen $\Lambda$ and the BHB$\Lambda \phi$ are adopted in 
these calculations. Under each EoS, the first column represents the black hole 
formation time ($t_{BH}$) estimated from the core bounce and the next column
gives the maximum baryon mass (M$_{b,max}$) followed by the maximum
gravitational mass (M$_{g,max}$) of the PNS at the point of instability 
corresponding to the central value of the lapse function 0.3. Further
investigations with the two $\Lambda$ hyperon EoSs reveal an opposite behaviour
of
$t_{BH}$ than what has been observed for nuclear EoSs. For $\Lambda$ 
hyperon EoS, $t_{BH}$ for the BHB$\Lambda \phi$ is always
greater than that of the HShen $\Lambda$ for all presupernova models except 
$s$40WH07. The comparison of two hyperon EoSs shows that the BHB$\Lambda\phi$ 
is
a stiffer EoS than the HShen $\Lambda$. The explanation of this behaviour may 
be traced back to the inclusion of repulsive $\Lambda$-$\Lambda$ interaction in
the BHB$\Lambda \phi$ EoS. For all presupernova models and EoSs adopted in 
simulations, it is evident from the table that the maximum gravitational mass 
of the PNSs in each case is higher than their corresponding maximum cold 
neutron star masses. However, in some cases, the maximum gravitational mass of 
the PNS collapsing
into a black hole with the HShen $\Lambda$ EoS is less than the  
two solar mass limit because the HShen $\Lambda$ EoS does not result in a 
2 M$_{\odot}$ cold neutron star. It is interesting to note that in the case of 
the HShen $\Lambda$ EoS, the difference 
between M$_{g,max}$ of the PNS and the maximum mass of the cold neutron star 
that includes $\Lambda$ hyperons (1.75 M$_{\odot}$) is appreciable, whereas the 
maximum gravitational mass of the PNS for the BHB$\Lambda \phi$ EoS is very 
similar to the value of the corresponding maximum mass of the cold neutron star
with $\Lambda$s (2.1 M$_{\odot}$) for the entire set of progenitor models. This 
shows that the thermal effects in the PNS for the BHBH$\Lambda \phi$ might 
not be as strong as in the PNS with the HShen $\Lambda$ because the EoS is 
stiffer in the former case. The role of 
decreasing thermal pressure with increasing stiffness of the EoS was already 
noted by \citet{ott11}. This should have interesting implications for the study
of the metastability of the PNS with the BHB$\Lambda \phi$ EoS.

We compare our findings with other CCSN simulations with hyperon EoS. The 
Ishizuka hyperon EoS includes $\Lambda$, $\Sigma$, and $\Xi$ hyperons and
is an extension of the HShen nuclear EoS \citep{ishi}. 
The CCSN simulations were performed in a spherically symmetric general
relativistic neutrino radiation hydrodynamics model using a 40 M$_{\odot}$ 
progenitor of \citet{weav} and the Ishizuka hyperon EoS \citep{sumi,naka}. 
With the LS+$\Lambda$ EoS \citep{oertel12} 
Peres et al. \citep{per} carried out a similar investigation 
using an $s$40WW progenitor and a low metallicity 40 M$_{\odot}$ progenitor 
of \citet{whw} called $u$40. 
\citet{sb1} also studied CCSN simulations using the HShen 
$\Lambda$ EoS and progenitor models of \citet{Woos}, particularly 
studying the long duration
evolution of the PNS in the context of understanding the fate of the compact 
object in SN1987A. It should be noted that though our results with the 
BHB$\Lambda \phi$ EoS are qualitatively similar to those of earlier 
calculations, they are quantitatively different because only our $\Lambda$
hyperon EoS is compatible with the 2 M$_{\odot}$ limit of cold neutron stars. 
The early black hole formation due to softening in the $\Lambda$ hyperon EoS 
compared with the nuclear EoS is a robust conclusion in all of these 
calculations. 
Total neutrino luminosity as well as $\nu_e$, $\bar{\nu}_e$, and $\nu_x$ 
luminosities as a function of postbounce time are plotted in Figure 8
for the HS(DD2) (left panel) and the BHB$\Lambda \phi$ (right panel) EoS. The
results are shown here for the $s$40WH07 model. It should be noted that the 
neutrino emission 
ceases earlier for the BHB$\Lambda \phi$ case than for the scenario with the 
HS(DD2) nuclear EoS. A similar conclusion was arrived at in the simulation with 
other hyperon EoSs \citep{sumi,sb1}. The shorter neutrino burst 
corresponding to the $\Lambda$ hyperon EoS before the collapse of the PNS into 
the black hole could be an important probe for the appearance of $\Lambda$ 
hyperons in the PNS. This demands a more accurate treatment of neutrinos in 
the GR1D code. Figure 9 exhibits the neutrino luminosities
for both the $\Lambda$ hyperon EoS and the $s$40WH07 model. We find similar
features for neutrino luminosities for both cases.  
Though we are considering a phase transition 
from nuclear to $\Lambda$ hyperon matter, we do not find any evidence for a 
second neutrino burst, which was observed in a first order quark-hadron phase 
and was responsible for a successful supernova explosion \citep{sag}. 

So far we have seen that simulations in the 1D CCSN model might lead 
to accretion driven black holes in failed supernovae. If a successful supernova
occurs, can exotic matter such as hyperons make the PNS metastable and drive
it to become a low mass black hole during the long duration evolution when 
thermal
support decreases and deleptonization takes place in the PNS? Such a scenario
was envisaged for the non-observation of a compact object in SN1987A 
\citep{bethe,cook,sb3}. This problem was also studied in CCSN simulations 
\citep{keil,baum,sb1}. 
We continue our study by increasing the neutrino heating scale factor to 
$f_{heat}=$ 1.5 for $s$20WH07 with the BHB$\Lambda \phi$ EoS. The left panel of 
Figure 10 exhibits 
the shock radius as a function of postbounce time. For the neutrino scale
factor $f_{heat} =$ 1, it fails to launch a successful supernova explosion and
the shock radius recedes. Finally, the PNS collapses into a black hole. For
$f_{heat} =$ 1.5, it is observed that the shock radius increases with time
after a successful supernova explosion. The PNS remains stable until 4 s.
We do not find any onset of the metastability in the PNS due to the loss of 
thermal support and neutrino pressure during the cooling phase over a few 
seconds. The window for the metastability is very narrow because the maximum
PNS mass in this case is 2.138 M$_{\odot}$ whereas the maximum cold neutron
star mass corresponding to the BHB$\Lambda \phi$ EoS is 2.1 M$_{\odot}$. The PNS
might evolve into a cold neutron star. 
Gravitational masses of the PNS for $f_{heat} =$ 1 and 1.5  are shown as a 
function of postbounce time in the right panel of Figure 10.  The PNS 
cools down to a neutron star with a mass $\sim 1.64$ M$_{\odot}$ at the end
of 4 s. 

\section{Summary and Conclusions}       

We have performed CCSN simulations using the BHB$\Lambda \phi$ EoS, which is 
compatible with a 2 M$_{\odot}$ neutron star, 
and several progenitor models from the stellar studies of \citet{Woos}. It is
observed that $\Lambda$s are produced a few hundred milliseconds after the 
core bounce. The appearance of $\Lambda$ hyperons is studied in great detail.
It is evident from the density and temperature profiles as a function of radius
that $\Lambda$s are produced in the core of the PNS when the central density
exceeds two times the normal nuclear matter density during the postbounce 
evolution
phase. On the other hand, an off-centre population of thermal $\Lambda$ 
hyperons 
is the result of peak values of temperature away from the centre of the PNS. 
When we set the neutrino heating scale factor $f_{heat}=$ 1, each CCSN 
simulation ends with the formation of a black hole driven by mass 
accretion. It is interesting to find out that the black hole formation time for
the BHB$\Lambda \phi$ EoS is shorter than that of the HShen $\Lambda$ EoS
though the opposite conclusion is drawn from the accretion driven black hole 
with the HShen nuclear and HS(DD2) EoS models. This is attributed to the fact 
that the repulsive $\Lambda$-$\Lambda$ interaction in the BHB$\Lambda \phi$ EoS
makes it a stiffer EoS than the HShen $\Lambda$ EoS. Neutrino luminosity is 
found to cease with the formation of a black hole earlier for the 
$\Lambda$ hyperon EoS than for the corresponding case with the nuclear EoS.   
We have studied the metastability of the PNS due to the 
BHB$\Lambda \phi$ EoS in the long duration evolution after a successful 
supernova explosion using the $s$20WH07 progenitor model with the increased 
neutrino 
heating scale factor of $f_{heat} =$ 1.5. 
In this case, we do not find any delayed collapse into the black hole due to 
the presence of $\Lambda$ hyperons in the PNS.
The PNS that has a mass $\sim 1.64$ 
M$_{\odot}$ remains stable until 4 s and might become a cold neutron star. 

\acknowledgments

The numerical calculations presented in this article have been performed in the 
blade server of the Astroparticle Physics and Cosmological Division, Saha 
Institute of Nuclear Physics. We acknowledge fruitful discussions of GR1D with 
Evan O'Connor. DB thanks the Alexander von Humboldt Foundation for 
support.

%%%%%%%%%%%%%%%%%%%%%%%%%%%%%%%%%%%%%%%%%%%%%%%%%%%%%%%%%%%%%%%%%%%%%%%%%%%
\clearpage

\begin{deluxetable}{rccccccc}
\tablecolumns{17}
\tablewidth{0pc}
\tablecaption{
The saturation properties of nuclear matter
such as saturation density ($n_0$), binding energy (BE), incompressibility (K),
symmetry energy (S), and slope coefficient of symmetry energy (L) are obtained 
using the DD2 and TM1 parameter are obtained using the DD2 and TM1 parameter 
sets.} 
\tablehead{
Parameter& $n_0$ &BE&K&S&L&M$_{max}$\\
Set & ($fm^{-3}$) &(MeV)&(MeV)& (MeV) &(MeV)&(M$_{\odot}$)}
\startdata

DD2 &0.1491&16.02&243& 31.67&55.04&2.42\\  
TM1 &0.1455&16.31&281& 36.95&110.99&2.18\\  

\hline
\enddata
\tablecomments{
Maximum masses of cold 
neutron stars without $\Lambda$ hyperons corresponding to the HS(DD2) and the 
HShen EoS are also mentioned here.} 
\label{table}
\end{deluxetable}
%%%%%%%%%%%%%%%%%%%%%%%%%%%%%%%%%%%%%%%%%%%%%%%%%%%%%%%%%%%%%%%%%%%%%%%%%%%
\clearpage

\begin{deluxetable}{rccccccccccccccc}
\tablecolumns{17}
\tablewidth{0pc}
\tablecaption{Black hole formation time, baryonic and gravitational masses of
PNSs for CCSN simulations with the progenitor models of \citet{Woos} and 
the BHB$\Lambda \phi$ and HShen $\Lambda$ EoS tables.} 
\tablehead{
Model& \multicolumn{3}{c}{BHB$\Lambda\phi$}  && \multicolumn{3}{c} {HShen ${\Lambda}$} \\
& $t_{BH}$ &$M_{b, max}$&$M_{g, max}$&& $t_{BH}$ &$M_{b, max}$&$M_{g, max}$\\
& $(s)$ &$(M_{\odot})$&$(M_{\odot})$&& $(s)$ &$(M_{\odot})$&$(M_{\odot})$}
\startdata

$s$20WH07&1.938&2.251&2.138 && 1.652&1.999&1.964\\  
$s$23WH07& 0.879&2.276&2.203 && 0.847&2.095&2.073\\   
$s$25WH07& 1.548&2.234&2.141 && 1.376&2.035&2.001\\ 
$s$30WH07& 2.942&2.243&2.113 && 2.258&1.967&1.929\\ 
$s$35WH07& 1.175&2.243&2.161 && 1.084&2.071&2.041\\
$s$40WH07& 0.555&2.250&2.210 && 0.565&2.129&2.118\\

\hline
\enddata
\tablecomments{For all cases considered here, $f_{heat} =$ 1.}
\label{table1}
\end{deluxetable}
%%%%%%%%%%%%%%%%%%%%%%%%%%%%%%%%%%%%%%%%%%%%%%%%%%%%%%%%%%%%%%%%%%%%%%%%%%%
\clearpage

\begin{figure}
\epsscale{1.00}
\plotone{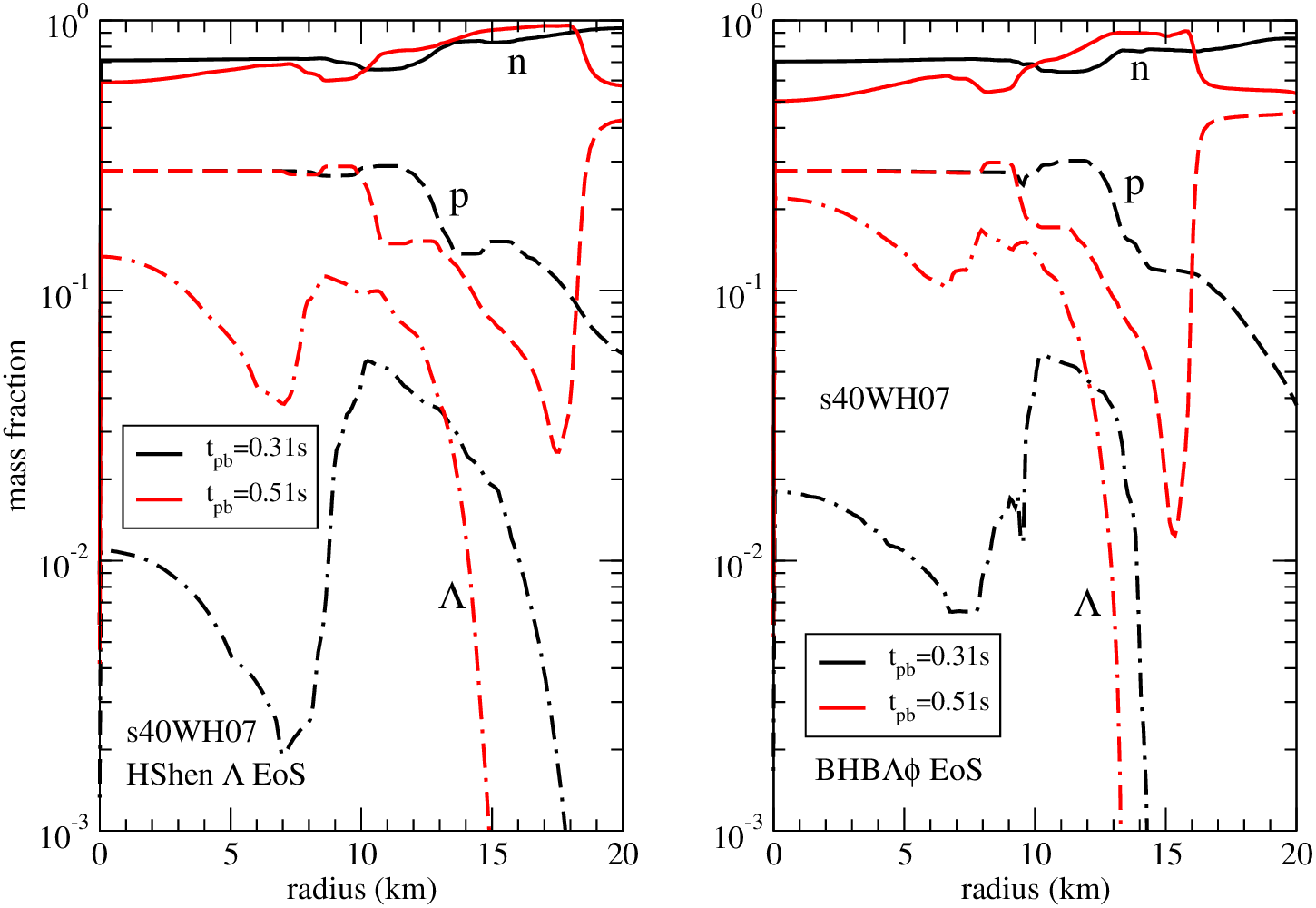}
\caption{Mass fractions of different species in the PNS are shown as a function
of radius for the HShen $\Lambda$ EoS (left panel) and the BHB$\Lambda \phi$ 
EoS (right panel) at $t_{pb} =$ 0.31 and 0.51 s (online-version: red).
The results in both panels correspond to the $s$40WH07 model.}
\end{figure}

\clearpage

\begin{figure}
\epsscale{.80}
\plotone{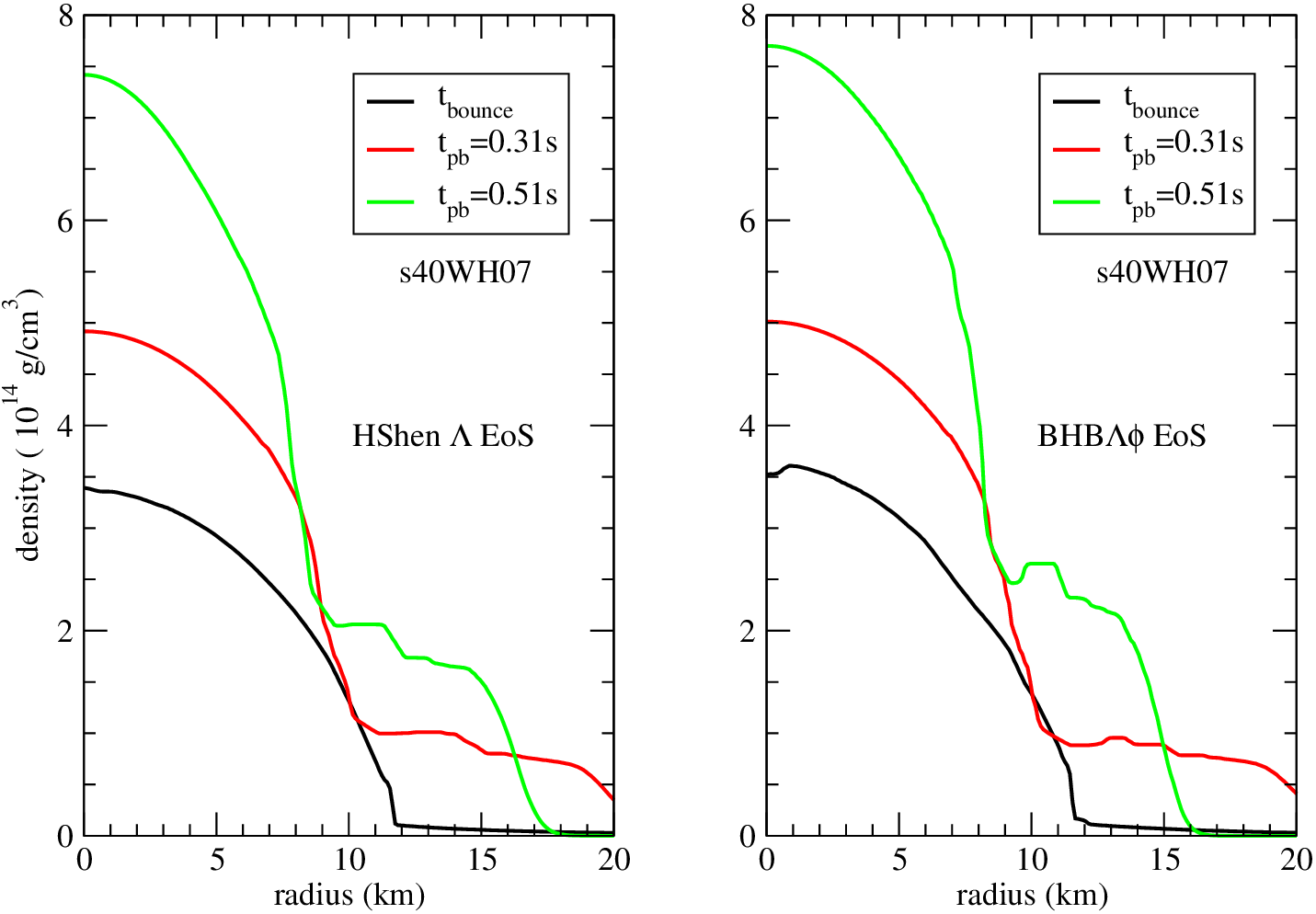}
\caption{Density profiles of the PNS are shown as a function of radius 
for the HShen $\Lambda$ EoS (left panel) and the BHB$\Lambda \phi$ 
EoS (right panel) 
at the core bounce and $t_{pb} =$ 0.31 (online-version: red) and 0.51 s 
(online-version: green). The results in both panels correspond to the $s$40WH07
model.}
\end{figure}

\clearpage

\begin{figure}
\epsscale{.80}
\plotone{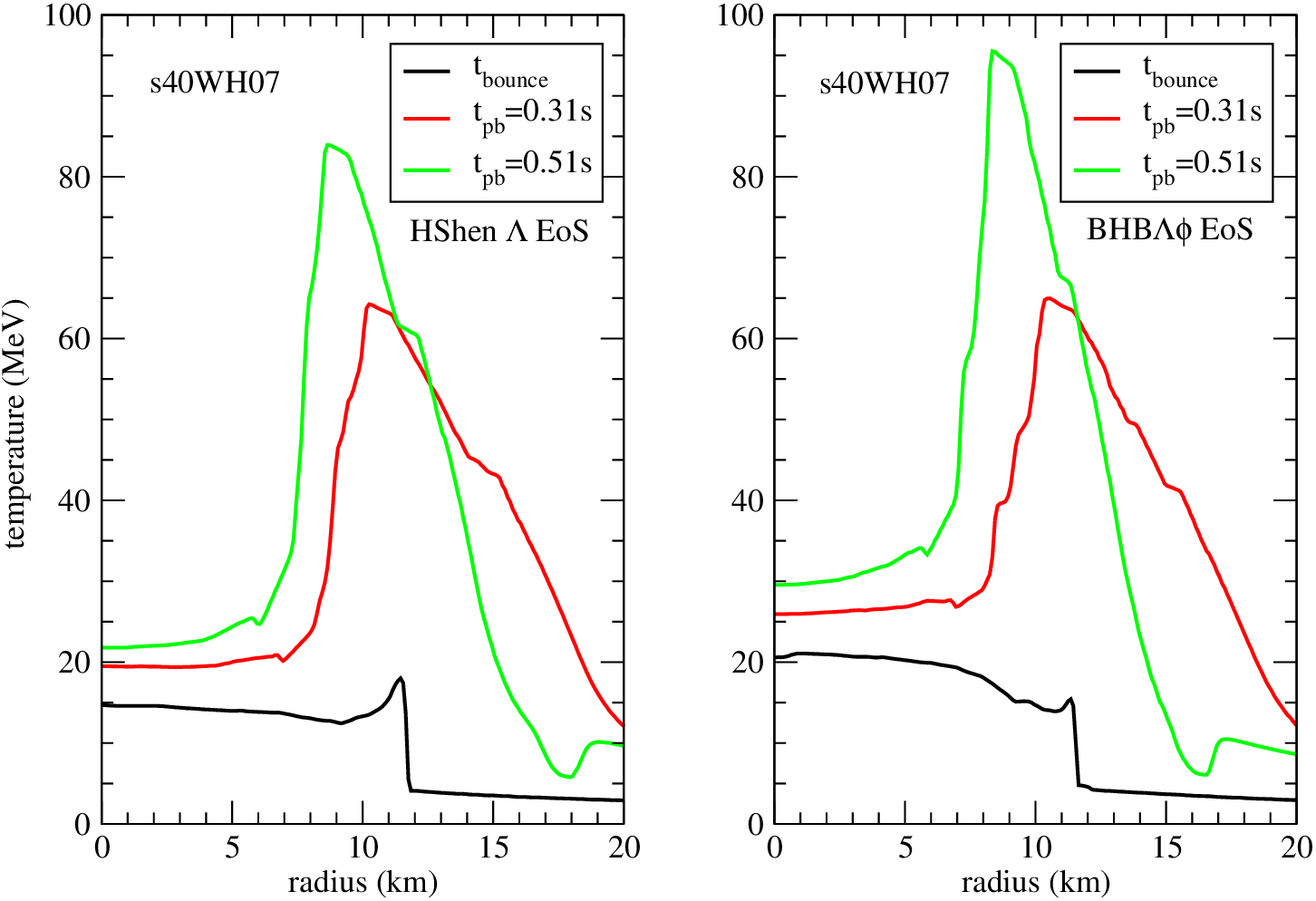}
\caption{Temperature profiles of the PNS are shown as a function of radius 
for the HShen $\Lambda$ EoS (left panel) and the BHB$\Lambda \phi$ 
EoS (right panel) 
at the core bounce and $t_{pb} =$ 0.31 (online-version: red) and 0.51 s 
(online-version: green). The results in both panels correspond to the $s$40WH07
model.}
\end{figure}

\clearpage

\begin{figure}
\epsscale{1.00}
\plotone{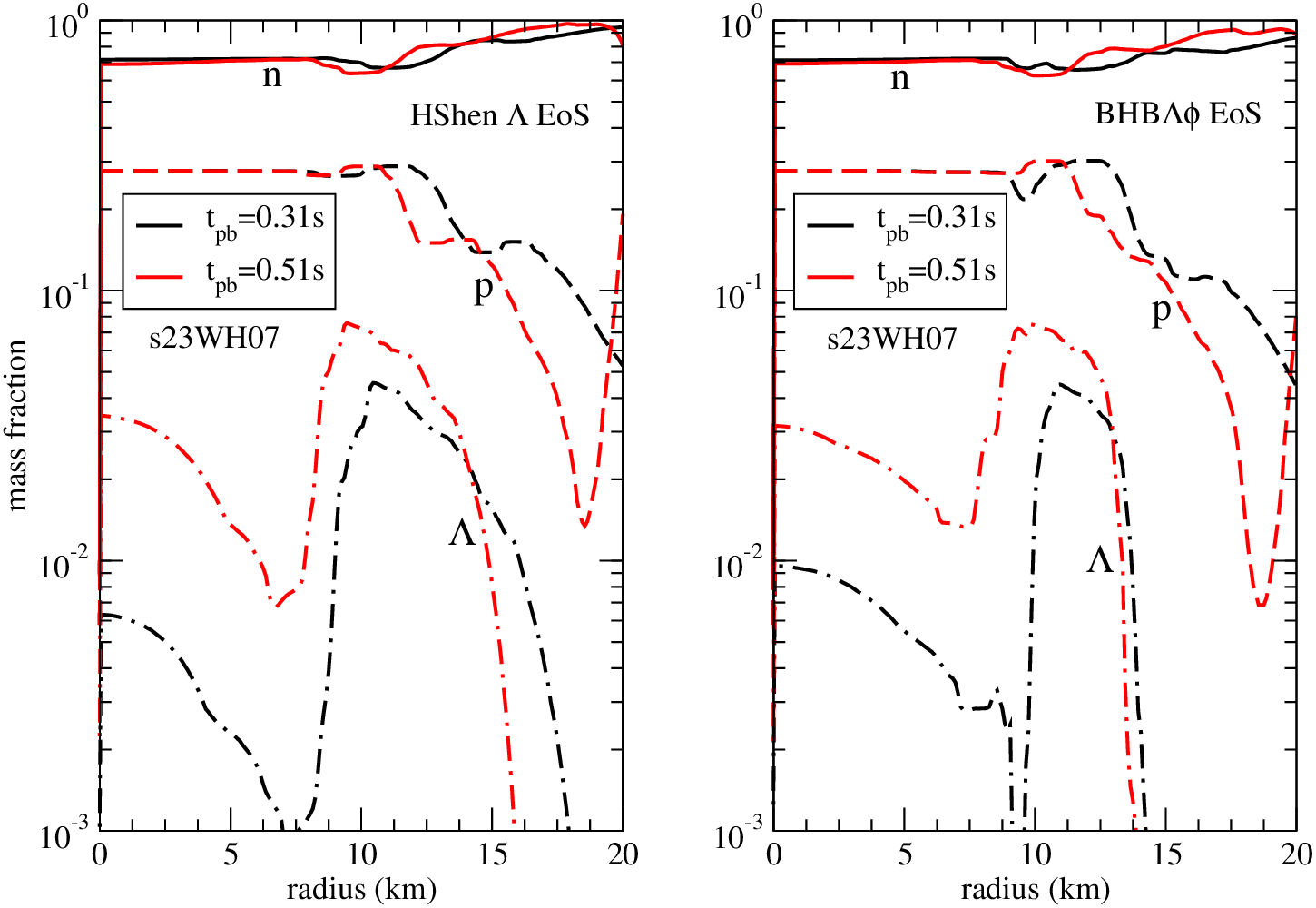}
\caption{Same as Figure 1 but for the $s$23WH07 model. The results correspond 
to the HShen $\Lambda$ EoS (left panel) and the BHB$\Lambda \phi$ EoS (right
panel) at $t_{pb} =$ 0.31 and 0.51 s (online-version: red).}
\end{figure}

\clearpage

\begin{figure}
\epsscale{.80}
\plotone{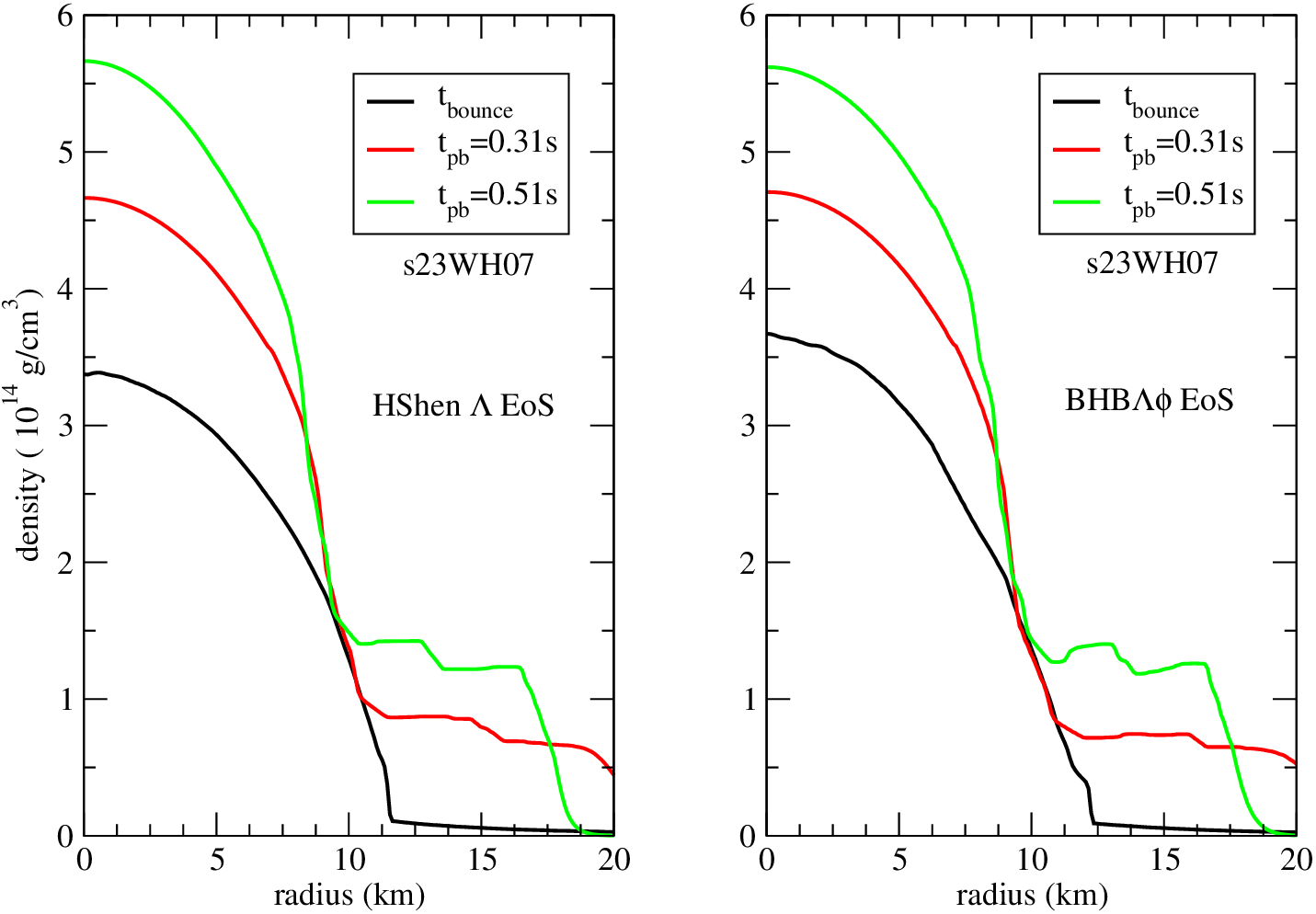}
\caption{Same as Figure 2 but for the $s$23WH07 model. The results correspond 
to the HShen $\Lambda$ EoS (left panel) and the BHB$\Lambda \phi$ EoS (right
panel) at the core bounce and $t_{pb} =$ 0.31 (online-version: red) and 
0.51 s (online-version: green).}
\end{figure}

\clearpage

\begin{figure}
\epsscale{.80}
\plotone{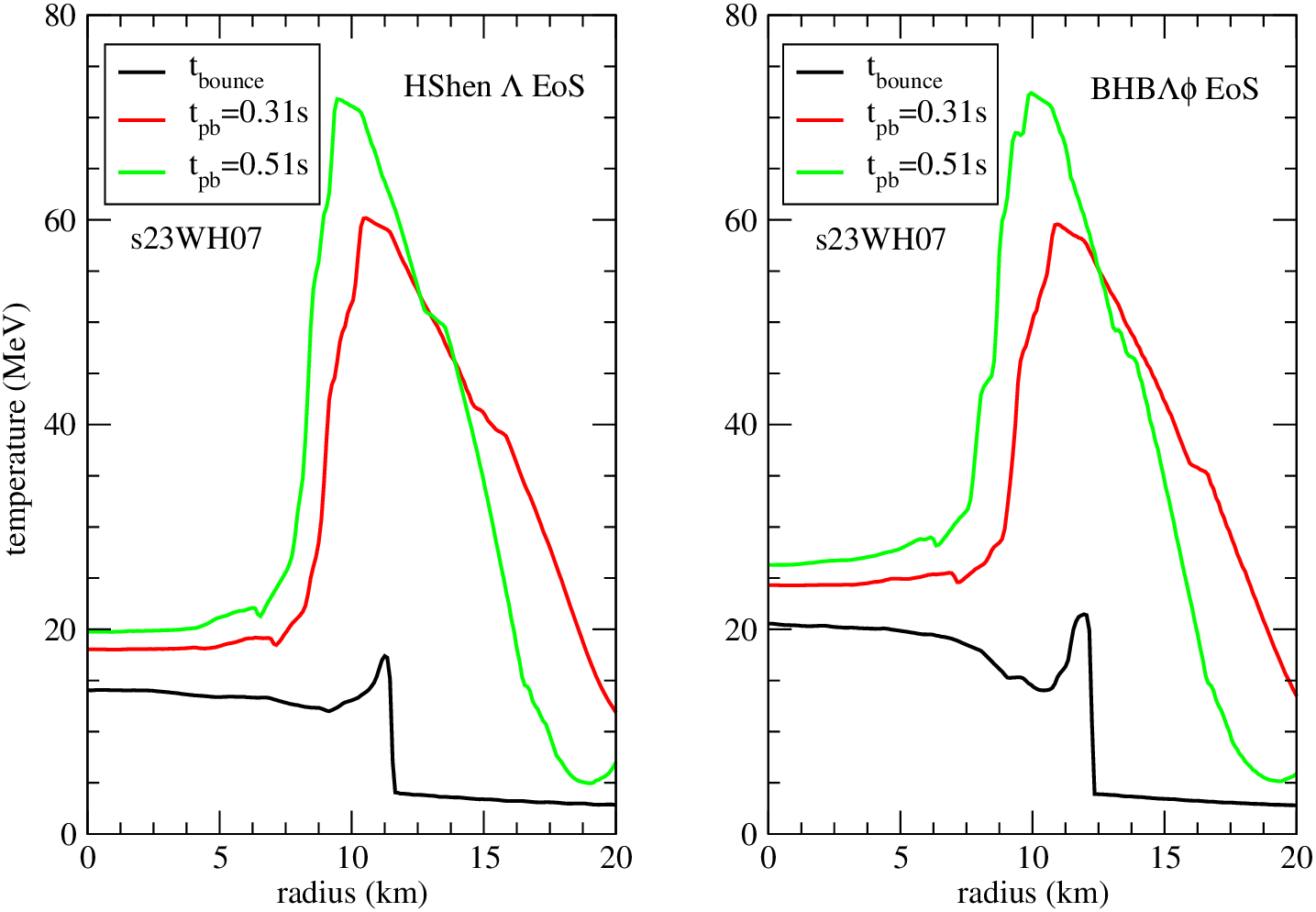}
\caption{Same as Figure 3 but for the $s$23WH07 model. The results correspond 
to the HShen $\Lambda$ EoS (left panel) and the BHB$\Lambda \phi$ EoS (right
panel) at the core bounce and $t_{pb} =$ 0.31 (online-version: red) and 
0.51 s (online-version: green).}
\end{figure}
\clearpage

\begin{figure}
\epsscale{1.00}
\plotone{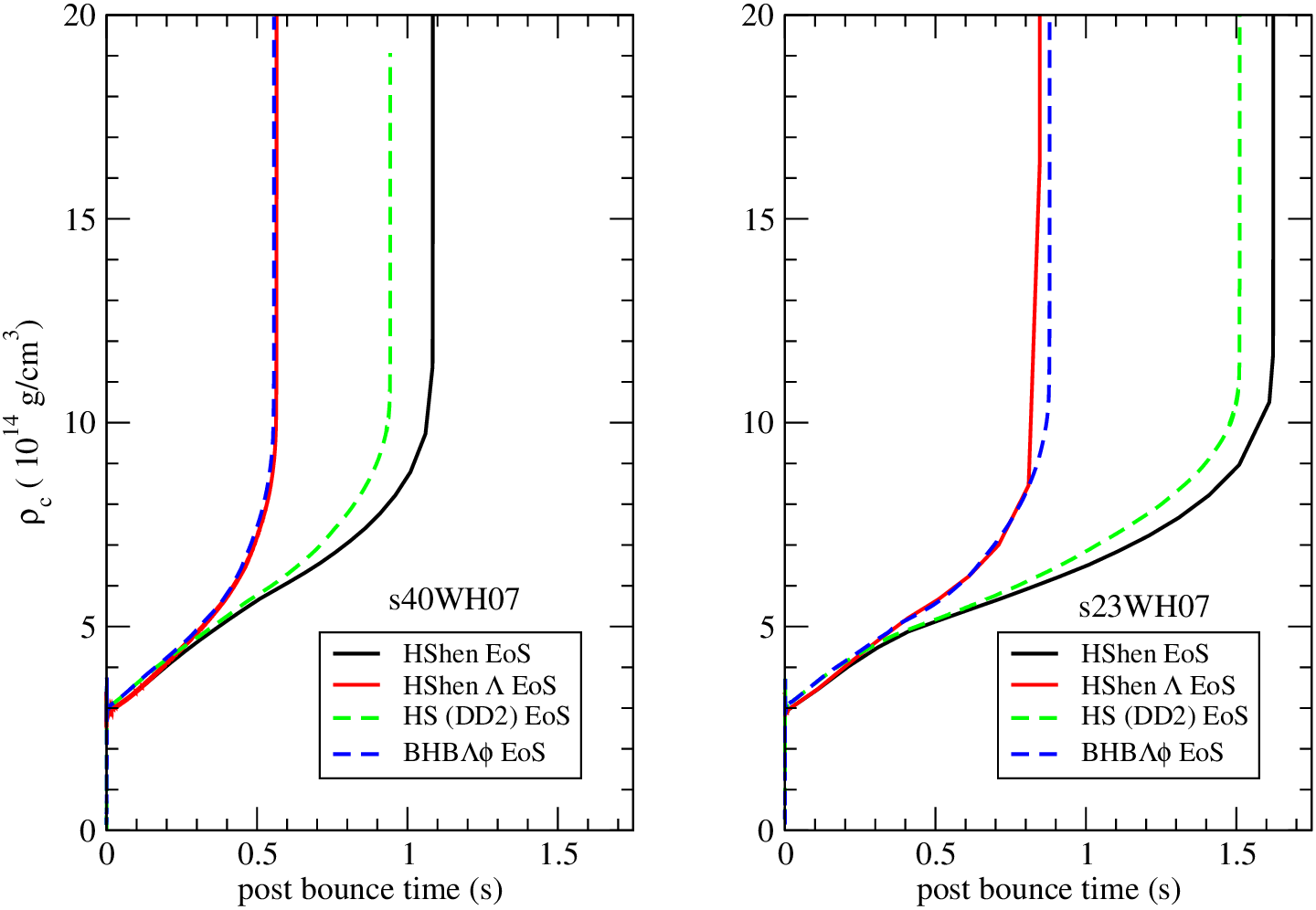}
\caption{Central baryon density is plotted with the postbounce time for
the HShen nuclear EoS, the HShen $\Lambda$ EoS (online-version: red), the 
HS(DD2) 
(online-version: green) and the BHB$\Lambda \phi$ EoS (online-version: blue). 
The results in the left and right panels correspond to the $s$40WH07 and 
$s$23WH07 models.}
\end{figure}

\clearpage

\begin{figure}
\epsscale{0.80}
\plotone{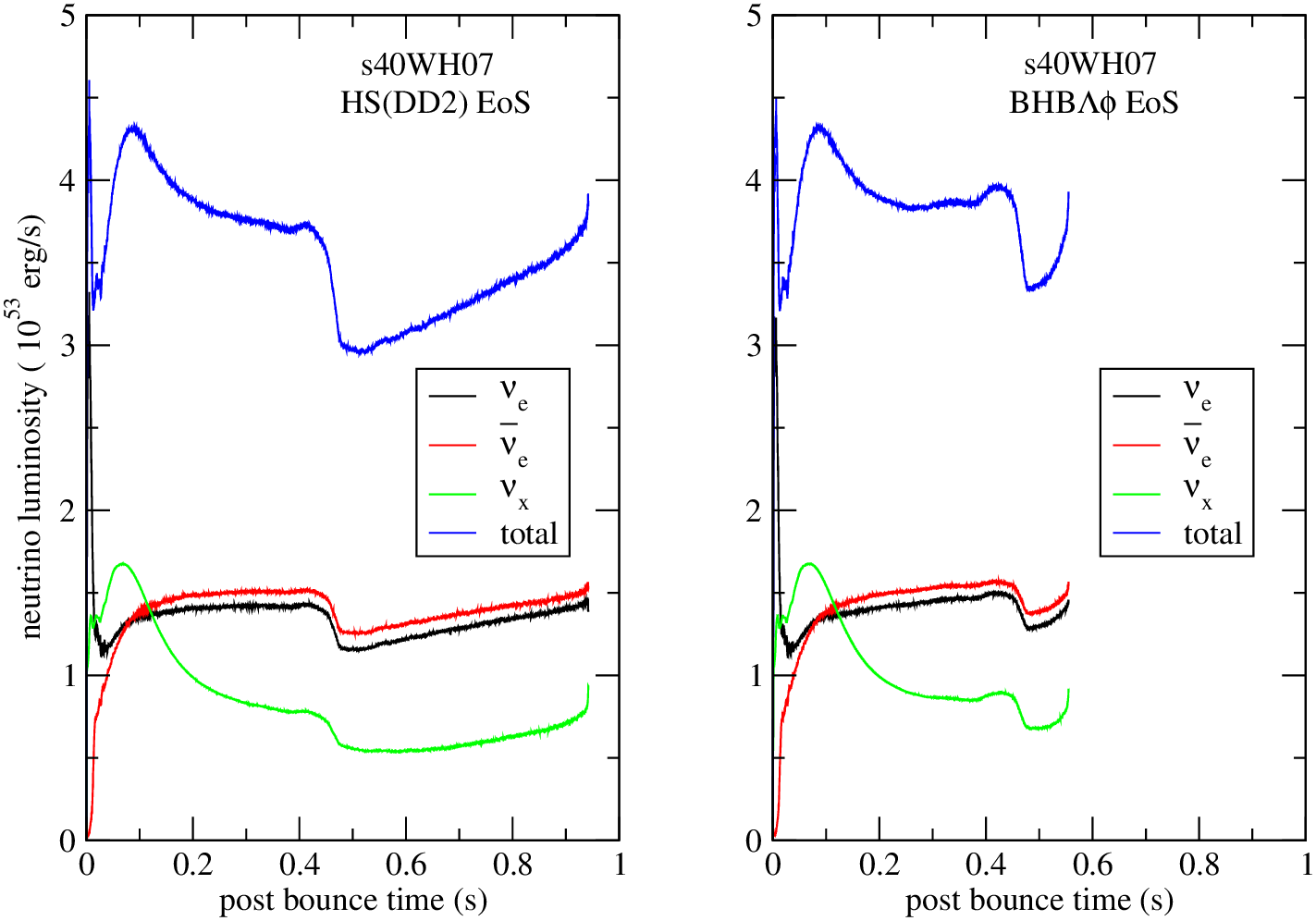}
\caption{Total neutrino luminosity (online-version: blue) as well as 
$\nu_e$, $\bar{\nu}_e$ (online-version: red) and $\nu_x$ 
(online-version: green) luminosities are plotted with the postbounce time for
the HS(DD2) (left panel) and the BHB$\Lambda \phi$ (right panel) EoS. The 
results correspond to the $s$40WH07 model.}
\end{figure}

\clearpage

\begin{figure}
\epsscale{0.80}
\plotone{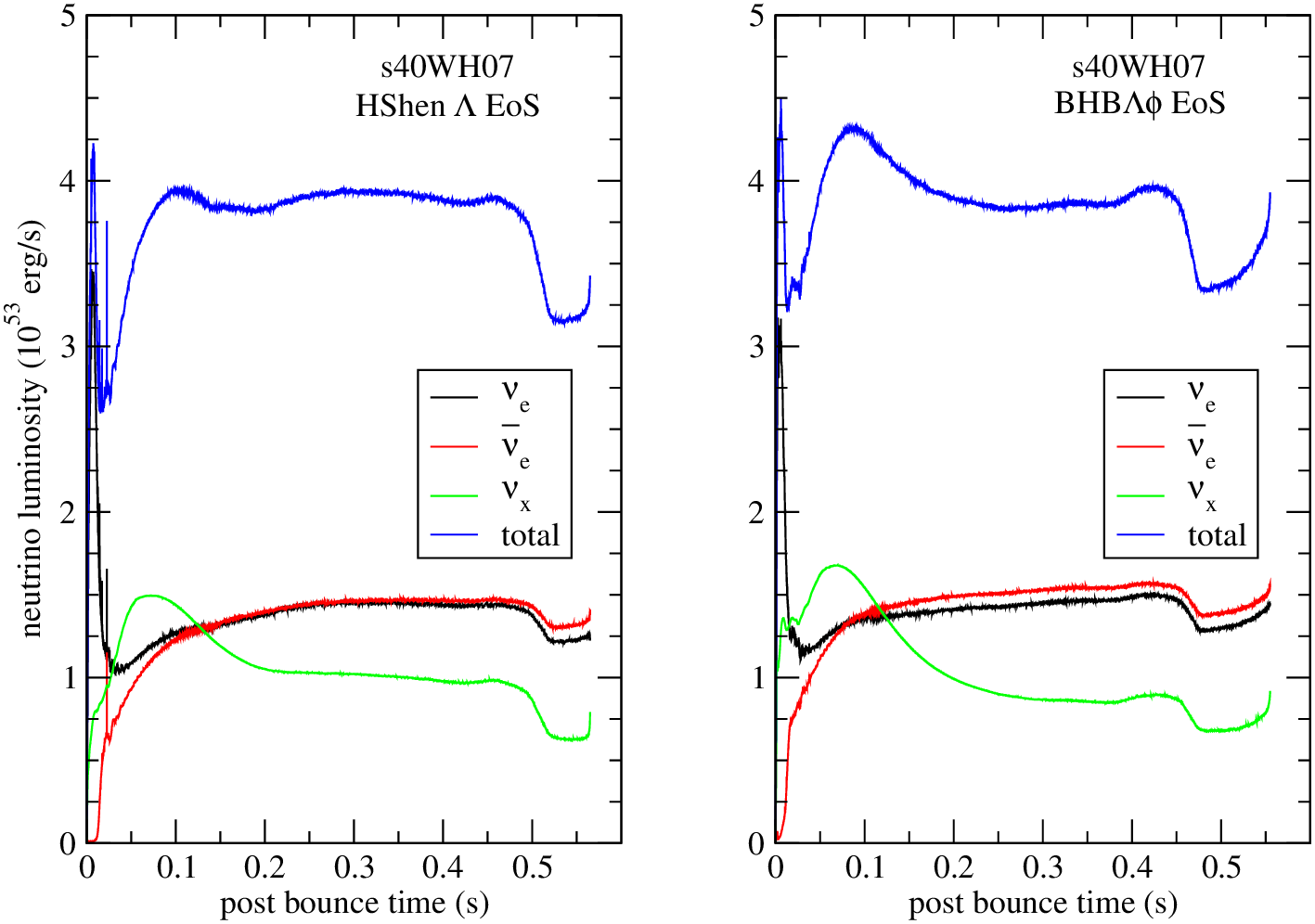}
\caption{Total neutrino luminosity (online-version: blue) as well as 
$\nu_e$, $\bar{\nu}_e$ (online-version: red) and $\nu_x$ 
(online-version: green) luminosities are plotted with the postbounce time for
the HShen $\Lambda$ (left panel) and the BHB$\Lambda \phi$ (right panel) EoS. 
The results correspond to the $s$40WH07 model.}
\end{figure}

\clearpage

\begin{figure}
\epsscale{1.00}
\plotone{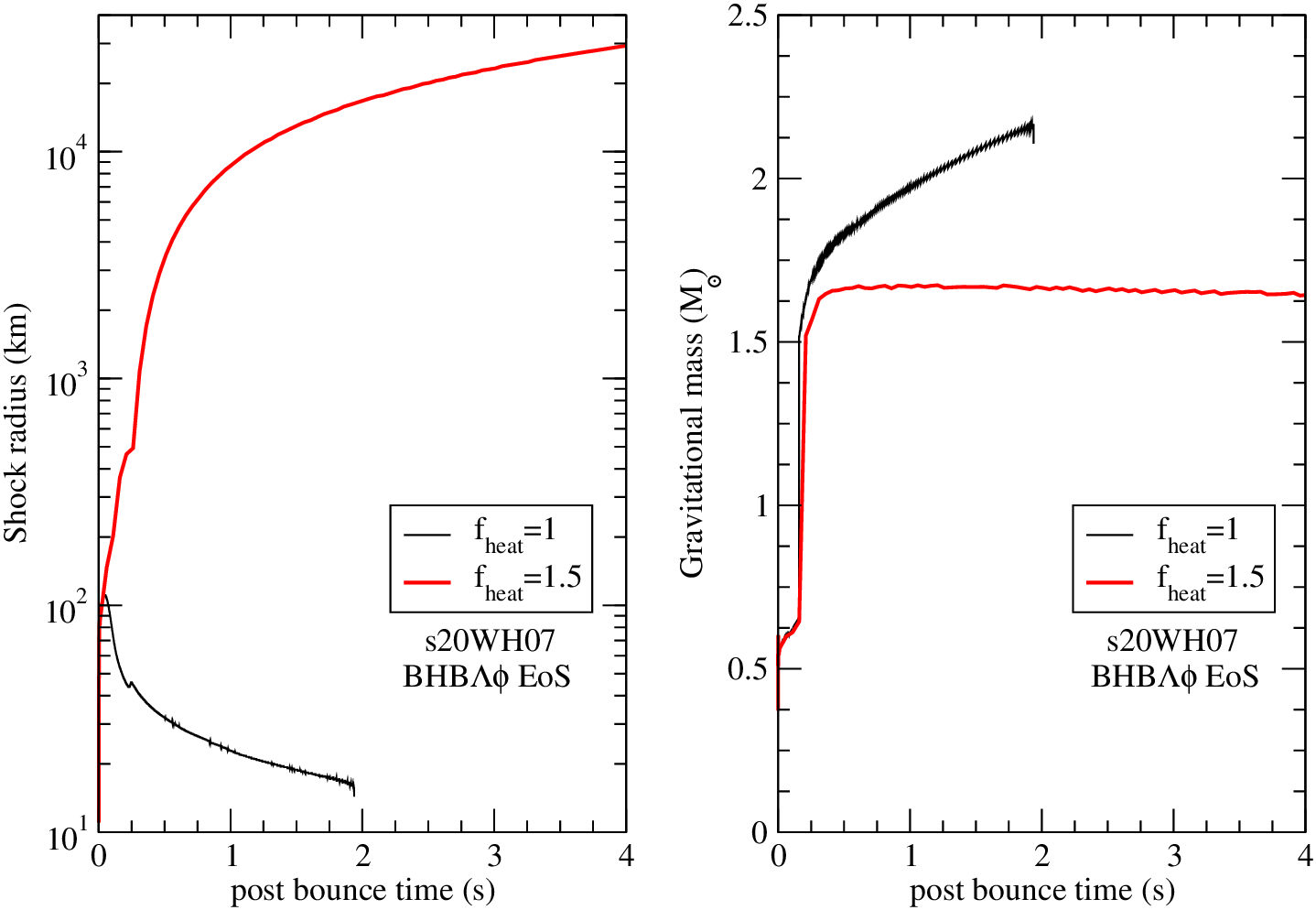}
\caption{Shock radius (left panel) and gravitational mass of the PNS (right 
panel) are plotted with the postbounce time using the neutrino heating 
factor $f_{heat}=$ 1 and 1.5 (online-version: red) for the $s$20WH07 model and 
the BHB$\Lambda \phi$ EoS.} 
\end{figure}

\end{document}